\documentclass{PoS}
\usepackage{epsfig}
\newcommand{\bea}{\begin{eqnarray}}
\newcommand{\eea}{\end{eqnarray}}
\newcommand{\be}{\begin{equation}}
\newcommand{\ee}{\end{equation}}

\newcommand{\as}{\alpha_s}
\newcommand{\asMZ}{\alpha_s(M^2_Z)}

\title{Strong coupling constant at NNLO from DIS data}

\ShortTitle{Strong coupling constant at NNLO from DIS data}

\author{\speaker{Anatoly Kotikov}%
         \thanks{
The work was supported in part by RFBR grant No.10-02-01259-a. 
The work of GP was supported in part by the grant Ministerio de Ciencia e 
Inovacion FPA2008-01177.
}\\
        Joint Institute for Nuclear Research, Russia\\
        E-mail: \email{kotikov@theor.jinr.ru}}

\author{Vasily~Krivokhizhin\\
        Joint Institute for Nuclear Research, Russia\\
        E-mail: \email{Vasily.Krivokhizin@lhep.jinr.ru}}

\author{Gonzalo~Parente\\
       Universidade de Santiago de Compostela, Spain\\
       E-mail: \email{gonzalo@fpaxp1.usc.es}}

\author{Binur~Shaikhatdenov\\
        Joint Institute for Nuclear Research, Russia\\
        E-mail: \email{sbg@theor.jinr.ru}}

\abstract{We discuss the results of
our recent analysis \cite{BKKP} of
deep inelastic scattering data on $F_2$ structure function
in the non-singlet approximation with next-to-next-to-leading-order accuracy.
The study of high statistics deep inelastic scattering data provided by BCDMS, SLAC,
NMC and BFP collaborations was performed with a special emphasis placed on 
the higher twist contributions.
For the coupling constant the following
value $\as(M_Z^2) = 0.1167 \pm 0.0022 
$ {\it (total exp. error)}
was found.
}

\FullConference{The XIXth International Workshop on High Energy Physics and
Quantum Field Theory \\
                 8-15 September 2010\\
                 Golitsyno, Moscow, Russia}

\begin{document}

\section{ Introduction }


It is already a common knowledge that the accuracy of data for DIS structure 
functions (SFs) allows one to study $Q^2$-dependence of logarithmic QCD-inspired corrections and those of power-like (non-perturbative) nature independently (see for instance~\cite{Beneke} and references therein). And this aspect is crucial for the analysis to be performed within some well defined scheme.


In this contribution we present the results of our recent analysis~\cite{BKKP}
of DIS SF $F_2(x,Q^2)$ carried out over SLAC, NMC, BCDMS and BFP experimental
data~\cite{SLAC1}
at NNLO of massless perturbative QCD.
%
As in our previous papers \cite{Kri,PKK,KK2001} the function $F_2(x,Q^2)$ 
is represented as a sum of the leading twist $F_2^{pQCD}(x,Q^2)$ and the 
twist four terms:
\be
F_2(x,Q^2)=F_2^{pQCD}(x,Q^2)\left(1+\frac{\tilde h_4(x)}{Q^2}\right)\,.
\label{1.1}
\ee

As is known there are at least two ways to perform QCD analysis over DIS 
data: the first one (see e.g.~\cite{ViMi,fits}) deals
with Dokshitzer-Gribov-Lipatov-Altarelli-Parisi (DGLAP) 
integro-differential equations~\cite{DGLAP} and let the data be examined 
directly, whereas the second one involves the SF moments and allows 
performing an analysis in analytic form as opposed to the former option.
In this work we take on the way in-between these two latter, i.e. analysis 
is carried out over the moments of SF $F_2^{k}(x,Q^2)$ defined as follows
\be
M_n^{pQCD/twist2/\ldots}(Q^2)=\int_0^1 x^{n-2}\,F_2^{pQCD/twist2/\ldots}(x,Q^2)\,dx
\label{1.a}
\ee
and then reconstruct SF for each $Q^2$ by using the  Jacobi polynomial 
expansion method~\cite{Kri,Barker}.
The theoretical input can be found
in the papers~\cite{KK2001,K2007}.

\section{ A fitting procedure }
\label{sec3}

The fitting procedure largely follows that used in~\cite{KK2001}.
With the QCD expressions for the Mellin moments $M_n^{k}(Q^2)$ analytically calculated according to
the formul\ae\, given above the SF $F_2^k(x,Q^2)$ is reconstructed by using the Jacobi polynomial expansion method:
$$
F_{2}^k(x,Q^2)=x^a(1-x)^b\sum_{n=0}^{N_{max}}\Theta_n ^{a,b}(x)\sum_{j=0}^{n}c_j^{(n)}(\alpha ,\beta )
M_{j+2}^k (Q^2)\,,
\label{2.1}
$$
where $\Theta_n^{a,b}$ are the Jacobi polynomials and $a,b$ are their parameters to be fitted. A condition
imposed on the latter is the requirement of the error minimization while reconstructing the structure functions.

Since a twist expansion starts to be applicable only above $Q^2 \sim 1$ GeV$^2$
the cut $Q^2 \geq 1$ GeV$^2$ is imposed on the experimental data throughout.
The MINUIT program~\cite{MINUIT} is used to minimize two variables
$$
\chi^2_{SF} = \biggl|\frac{F_2^{exp} - F_2^{teor}}{\Delta F_2^{exp}}\biggr|^2\,, \qquad
\chi^2_{slope} = \biggl|\frac{D^{exp} - D^{teor}}{\Delta D^{exp}}\biggr|^2\,,
$$
where $D=d\ln F_2/d\ln\ln Q^2$. The quality of the fits is characterized by
$\chi^2/{\rm DOF}$ for the SF $F_2$. Analysis is also performed for the 
slope $D$ that serves the purpose of checking the properties of 
fits.

We use free normalizations of the data for different experiments.
For a reference set, the most stable deuterium BCDMS data at the value of the
beam initial energy $E_0=200$ GeV is used.

\section{Results}

Since the gluon distribution function is not taken into account in the nonsinglet approximation, the analysis is substantially
easier to conduct; hence the cut on the Bjorken variable ($x\geq 0.25$) imposed
where gluon density is believed to be negligible.
The starting point of the evolution is taken to be
$Q^2_0$ = 90 GeV$^2$.
These $Q^2_0$ values are close to the average values of $Q^2$ spanning the corresponding data.
The previous experience tells us that the maximal value of the number
of moments to be accounted for is $N_{max} =8$~\cite{Kri} (though we
check $N_{max}$ dependence just like in the NLO analysis) and
the cut $0.25 \leq x \leq 0.8$ is imposed everywhere.

In~\cite{KK2001,BKKP} the cuts on the kinematic variable $Y=(E_0-E)/E_0$
have been imposed so as to exclude BCDMS data with large systematic errors. Here
$E_0$ and $E$ are lepton initial and final energies, 
respectively.
Upon excluding the set of data with large systematic errors
considerably higher values of $\alpha_s(M^2_Z)$ are obtained and rather mild
dependence of its values on the choice of $Y$ cut is observed. 
For more details we refer to~\cite{BKKP,KK2001}.
Once these cuts are applied, a full set of data consists of 797 points.

\begin{table}
\begin{center}
\begin{tabular}{|l|c|c|c|c|c|c|}
\hline
& &  & &  &\\
$Q^2_{min}$ & $N$ of & HTC &$\chi^2(F_2)$/DOF &
$\as(90~\mbox{GeV}^2)$ $\pm$ stat & $\asMZ$ \\
& points &  &  & &  \\
\hline \hline
1.0 & 797 &  No & 2.20 & 0.1767 $\pm$ 0.0008 & 0.1164 \\
2.0 & 772 &  No & 1.14 & 0.1760 $\pm$ 0.0007 & 0.1162 \\
3.0 & 745 &  No & 0.97 & 0.1788 $\pm$ 0.0008 & 0.1173 \\
4.0 & 723 &  No & 0.92 & 0.1789 $\pm$ 0.0009 & 0.1174 \\
5.0 & 703 &  No & 0.92 & 0.1793 $\pm$ 0.0010 & 0.1176 \\
6.0 & 677 &  No & 0.92 & 0.1793 $\pm$ 0.0012 & 0.1176 \\
7.0 & 650 &  No & 0.92 & 0.1782 $\pm$ 0.0015 & 0.1171 \\
8.0 & 632 &  No & 0.93 & 0.1773 $\pm$ 0.0018 & 0.1167 \\
9.0 & 613 &  No & 0.93 & 0.1764 $\pm$ 0.0022 & 0.1163 \\
\hline \hline
1.0 & 797 & Yes & 0.98 & 0.1772 $\pm$ 0.0027 & 0.1167 \\
\hline
\end{tabular}
\end{center}
\caption{$\asMZ$ and $\chi^2$ in the case of the combined analysis (HTCs 
stands for higher twist corrections).}
\label{tab1}
\end{table}

To verify a range of applicability of perturbative QCD
we start with analyzing the data without a contribution of twist-four terms
(which means $F_2 = F_2^{pQCD}$) and perform several fits with the cut
$Q^2 \geq Q^2_{min}$ gradually increased.
From Table~1 it is seen that unlike the NLO analysis
the quality of the fits starts to appear fairly good from $Q^2=3$ GeV$^2$ 
onwards (at NLO, it starts at $Q^2=10$ GeV$^2$~\cite{KK2001}) .
Then, the twist-four corrections are added and the data with the usual
cut $Q^2 \geq 1$ GeV$^2$ imposed upon is fitted. It is clearly seen that as in
the NLO case (see~\cite{KK2001})
here the higher twists do sizably improve the quality of the fit,
with insignificant discrepancy in the values of the coupling constant to be quoted below.

\begin{table}
\begin{center}
\begin{tabular}{|l||c|c|}
\hline
& &   \\
$x$ & $\tilde h_4(x)$ for $H_2$ $\pm$ stat &  $\tilde h_4(x)$ for $D_2$ $\pm$ stat \\
\hline \hline
0.275& -0.183 $\pm$ 0.020  & -0.197 $\pm$ 0.009 \\
0.35 & -0.149 $\pm$ 0.028  & -0.171 $\pm$ 0.015 \\
0.45 & -0.182 $\pm$ 0.029  & -0.033 $\pm$ 0.031 \\
0.55 & -0.236 $\pm$ 0.052  &  0.142 $\pm$ 0.057 \\
0.65 & -0.180 $\pm$ 0.135  &  0.295 $\pm$ 0.108 \\
0.75 & -0.177 $\pm$ 0.182  &  0.303 $\pm$ 0.158 \\
\hline
\end{tabular}
\end{center}
\caption{HTC parameter values obtained in NNLO analysis.}
\label{tab2}
\end{table}

The parameter values of the twist-four term are presented in Table~2.
Note that these for $H_2$ and $D_2$ targets are obtained in separate fits by analyzing SLAC, NMC and BCDMS datasets taken together.
%
For illustrative purposes we visualize those for the hydrogen data in Fig.~1, 
where the HTCs
obtained at NLO and NNLO levels 
are seen to be compatible with each other within errors.
\begin{figure}
\includegraphics[width=0.7\textwidth]{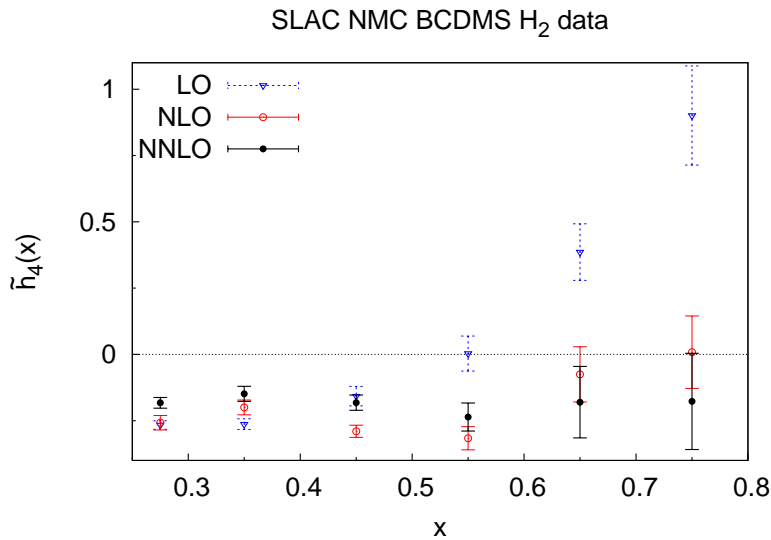}
\caption{ Comparison of the HTC parameter $\tilde h_4(x)$ obtained at LO, NLO and NNLO for hydrogen data (the bars indicate statistical errors).}
\label{fig1}
\end{figure}

We would like to note that the cut of the BCDMS data, which has increased the
$\as$ values (see Fig.~1 in~\cite{BKKP}) improves considerably agreement between perturbative QCD and experimental data.
Indeed, the HTCs, that are nothing else but the difference between
the twist-two approximation (i.e. pure perturbative QCD contribution) and the
experimental data, are seen to become considerably smaller at NLO and NNLO 
levels as compared with both the NLO higher twist terms obtained in~\cite{ViMi} and and the results of analysis obtained 
with no $Y$-cuts imposed on the BCDMS data (see Fig.~2).

\begin{figure}
\includegraphics[width=0.7\textwidth]{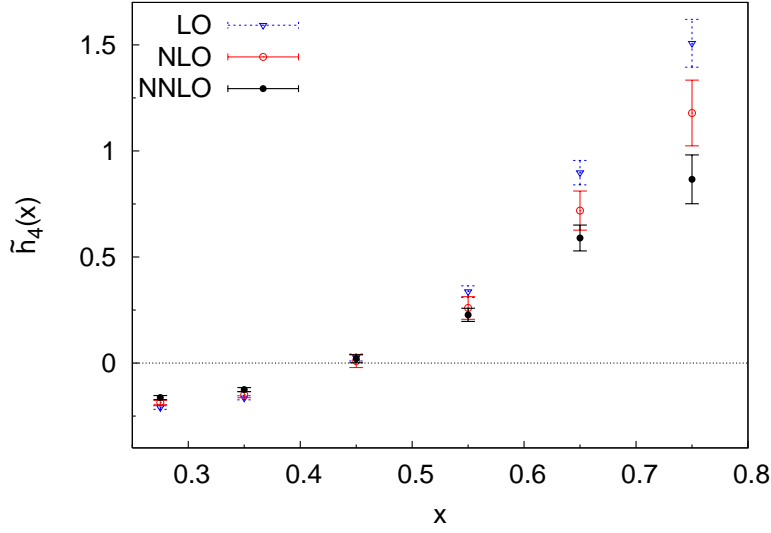}
\caption{ Comparison of the HTC parameter $\tilde h_4(x)$ obtained at LO, 
NLO and NNLO for hydrogen data 
when no $Y$ cuts imposed on the BCDMS data.
}
\label{fig2}
\end{figure}

\section{Conclusions}

In the paper \cite{BKKP} the Jacobi polynomial expansion method developed 
in~\cite{Kri,Barker}
was used to perform analysis of $Q^2$-evolution of DIS structure function $F_2$
by fitting all the existing to date reliable fixed-target experimental data that satisfy the cut $x \geq 0.25$.
Based on the results of fitting, the QCD coupling constant value at the normalization point was evaluated.
Starting with the reanalysis of BCDMS data by cutting off the points with large 
systematic errors it was shown \cite{BKKP,KK2001}
that the values of $\asMZ$ rise sharply with the cuts on systematics imposed.
The values of $\asMZ$ obtained in various fits are in agreement with each other.
An outcome is that quite a similar result for $\asMZ$ was obtained
\cite{BKKP} in the analysis performed over BCDMS (with the cuts on systematics) and the rest of the data, thus permitting us to fit available data altogether.

It turns out that for $Q^2 \geq 3$ GeV$^2$ the formulae of pure perturbative
QCD (i.e. twist-two approximation accompanied by the target mass corrections)
are enough to achieve good agreement with all the data analyzed.
The reference result is then found to be
\bea
\as(M_Z^2) &=& 0.1167 \pm 0.0008 ~\mbox{(stat)}
\pm 0.0018 ~\mbox{(syst)} \pm 0.0007 ~\mbox{(norm)} \nonumber \\
&=& 0.1167 \pm 0.0021~\mbox{(total exp. error)}\,.
 \label{re1n} 
\eea

Upon adding twist-four corrections, QCD (i.e. first two coefficients of Wilson expansion)
and the data are shown to be consistent with each other 
already at $Q^2 = 1$ GeV$^2$, where the Wilson
expansion begins to be applicable.
This way we obtain for the coupling constant at $Z$ mass peak:
\bea
\as(M_Z^2) &=& 0.1167 \pm 0.0007~\mbox{(stat)}
\pm 0.0020~\mbox{(syst)} \pm 0.0005~\mbox{(norm)} \nonumber \\
&=& 0.1167 \pm 0.0022 ~\mbox{(total exp. error)} \,.
\label{re2n} 
\eea

Note that the above values (\ref{re1n}) and (\ref{re2n}) are to some extent stable
\cite{CIKK09} under the application of the ``frozen'' \cite{frozen} and analytic \cite{SoShi}
modifications of the strong coupling constant, which as a rule lead to similar 
results (see~\cite{Zotov}).

Note also that our
results (\ref{re1n}) and (\ref{re2n}) for $\as(M_Z^2)$ are in good
agreement 
with the world average value for the
coupling constant presented in the review ~\cite{Breview},
\footnote{It should be mentioned that this analysis was carried out over
the data coming from the various experiments and in different orders of
perturbation theory, i.e. from NLO up to N$^{3}$LO.}
\be
\as(M_Z^2) = 0.1184 \pm 0.0007.
\label{re2n1}
\ee

Concerning the contributions of higher twist corrections in the present work the well-known $x$-shape of the twist-four corrections while going from intermediate to large values of the Bjorken variable $x$ is well reproduced.

\vskip 0.3cm
A.K. is indebted to organizers for the possibility
to present the talk which this paper is based on.

\end{document}